\pdfoutput=1 
\documentclass[twocolumn,aps,pra,superscriptaddress, floatfix]{revtex4-1}
\usepackage[utf8]{inputenc}
\usepackage{graphicx}
\usepackage{natbib}
\usepackage{url}
\usepackage{textcase}
\usepackage{bm}
\usepackage{color}
\usepackage{times}
\usepackage{amsmath}
\usepackage{physics}
\begin{document}
\title{A single-hole spin qubit}
\author{N.W. Hendrickx}
\email{n.w.hendrickx@tudelft.nl}
\author{W.I.L. Lawrie}
\author{L. Petit}
\affiliation{QuTech and Kavli Institute of Nanoscience, Delft University of Technology, P.O. Box 5046, 2600 GA Delft, The Netherlands}
\author{A. Sammak}
\affiliation{QuTech and Netherlands Organisation for Applied Scientific Research (TNO), Stieltjesweg 1, 2628 CK Delft, The Netherlands}
\author{G. Scappucci}
\author{M. Veldhorst}
\email{m.veldhorst@tudelft.nl}
\affiliation{QuTech and Kavli Institute of Nanoscience, Delft University of Technology, P.O. Box 5046, 2600 GA Delft, The Netherlands}
\date{\today}

\maketitle
\textbf{
Qubits based on quantum dots have excellent prospects for scalable quantum technology due to their inherent compatibility with standard semiconductor manufacturing \cite{vandersypen_interfacing_2017}. While early on it was recognized that holes may offer a multitude of favourable properties for fast and scalable quantum control \cite{bulaev_electric_2007, bulaev_spin_2005}, research thus far has remained almost exclusively restricted to the simpler electron system \cite{petta_coherent_2005,veldhorst_two-qubit_2015, huang_fidelity_2019}. However, recent developments with holes have led to separate demonstrations of single-shot readout \cite{vukusic_single-shot_2018} and fast quantum logic \cite{maurand_cmos_2016, watzinger_germanium_2018, hendrickx_fast_2019}, albeit only in the multi-hole regime. Here, we establish a single-hole spin qubit in germanium and demonstrate the integration of single-shot readout and quantum control. Moreover, we make use of Pauli spin blockade, allowing to arbitrarily set the qubit resonance frequency, while providing large readout windows. We deplete a planar germanium double quantum dot to the last hole, confirmed by radio-frequency reflectrometry charge sensing, and achieve single-shot spin readout. To demonstrate the integration of the readout and qubit operation, we show Rabi driving on both qubits and find remarkable electric control over their resonance frequencies. Finally, we analyse the spin relaxation time, which we find to exceed one millisecond, setting the benchmark for hole-based spin qubits. The ability to coherently manipulate a single hole spin underpins the quality of strained germanium and defines an excellent starting point for the construction of novel quantum hardware.
}

Group-IV semiconductor spin qubits \cite{loss_quantum_1998} are promising candidates to form the main building block of a quantum computer due to their high potential for scalability towards large 2D-arrays \cite{lawrie_quantum_2019,vandersypen_interfacing_2017,veldhorst_silicon_2017,li_crossbar_2018} and the abundance of net-zero nuclear spin isotopes for long quantum coherence \cite{itoh_isotope_2014, veldhorst_addressable_2014}. Over the past decade, all prerequisites for quantum computation were demonstrated on electron spin qubits in silicon, such as single-shot readout of a single electron \cite{morello_single-shot_2010}, high-fidelity single-qubit gates 
\cite{yoneda_quantum-dot_2017,yang_silicon_2019} and the operation of a two-qubit gate \cite{veldhorst_two-qubit_2015,huang_fidelity_2019, watson_programmable_2018,zajac_resonantly_2018}. 
However, hole spins may offer several advantages, such as a strong spin-orbit coupling (SOC) and a large excited state energy. Early research demonstrated the feasibility of using the SOC for all-electric driving \cite{pribiag_electrical_2013, nowack_coherent_2007}, but these experiments were limited by nuclear spins and reaching the single-hole regime remained an open challenge. More recently, hole spins in group-IV materials have gained attention as a platform for quantum information processing \cite{maurand_cmos_2016, watzinger_germanium_2018, hendrickx_fast_2019}. In particular hole states in germanium can provide a high degree of quantum dot tunability \cite{hendrickx_gate-controlled_2018, hardy_single_2019, froning_single_2018}, fast and all-electrical driving \cite{hendrickx_fast_2019, watzinger_germanium_2018} and ohmic contacts to superconductors for hybrids \cite{hendrickx_ballistic_2019,vigneau_germanium_2019}. These experiments culminated in the recent demonstration of full two-qubit logic \cite{hendrickx_fast_2019}. While hole spins have been read out in single-shot mode using the Elzerman technique \cite{vukusic_single-shot_2018}, these experiments require magnetic fields impractical for hole qubit operation due to the strongly anisotropic $g$-factor of hole spins in germanium \cite{mizokuchi_ballistic_2018}. Pauli spin blockade readout allows for spin readout independent of the Zeeman splitting of the qubit, leveraging the large excited state energy purely defined by the orbital energy for holes in germanium. Furthermore, achieving these assets on a single-hole spin demonstrates full control over the materials system and allows to tune the quantum dot occupancy at will, optimizing the different qubit properties. Moreover, the ability to study a platform at the single-particle level would provide great insight into its physical nature, crucial for holes which originate from a more complicated band structure than electrons \cite{he_electronic_2005,liles_spin_2018}.

In this work, we make this step and demonstrate for the first time single-shot readout and operation of a single hole spin qubit. We grow undoped strained germanium quantum wells \cite{sammak_shallow_2019} and fabricate devices using standard manufacturing techniques \cite{lawrie_quantum_2019}. The high mobility and low effective mass \cite{lodari_light_2019} allow us to define quantum dots of relatively large size, alleviating the restraints on fabrication. We deplete the quantum dots to their last hole, confirmed by charge sensing using a nearby single hole transistor (SHT). The use of radio-frequency (RF) reflectometry \cite{schoelkopf_radio-frequency_1998,reilly_fast_2007, barthel_fast_2010} enables a good discrimination of the charge state, while maintaining a high measurement bandwidth to allow for fast spin readout. We make use of Pauli spin blockade to perform the spin-to-charge conversion \cite{ono_current_2002}, maximally taking advantage of the large excited state energy splitting of $E_\text{ST}=0.85$ meV and obtain single-shot spin readout. Finally, we demonstrate the integration of readout and qubit operation by performing all-electrically driven Rabi rotations on both qubits. Studying the control of a single hole qubit, we find a remarkably strong dependence of the resonance frequency on electric field and show a tunability of almost 1 GHz using only small electric potential variations. 

\begin{figure}
    \centering
	\includegraphics[width=\columnwidth]{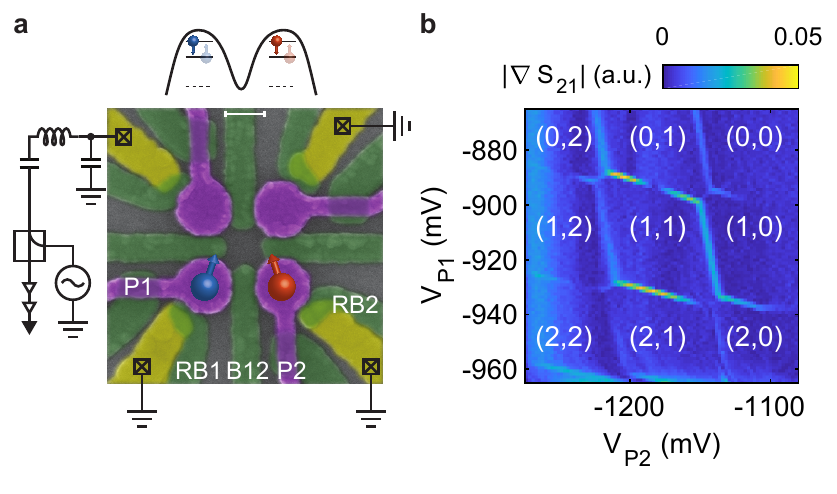}%
	\caption{
	\textbf{Fabrication and operation of a planar germanium double quantum dot.}
	\textbf{a} False-coloured scanning electron microscope image of the quadruple quantum dot device. We use the double quantum dot in the top channel as a single hole transitor (SHT) to sense changes in the charge occupation of the quantum dots formed under plunger gates P1 and P2. The charge sensor impedance is measured using reflectometry on a resonant circuit consisting of a superconducting inductor and the parasitic device capacitance. Barrier gates RB1 and RB2 can be used to control the tunnel rate of each quantum dot to its respective reservoir and gate B12 controls the interdot tunnel coupling.
	\textbf{b} Charge stability diagram of the double quantum dot system, where depletion of both quantum dots up to the last hole can be observed.
	\label{fig:1}}
\end{figure}

A false-coloured SEM picture of the quantum dot device is depicted in Figure \ref{fig:1}a. The device consists of a quadruple quantum dot system in a two-by-two array \cite{lawrie_quantum_2019}. We tune the top two quantum dots into the many-hole regime, such that they can be operated as a single hole transistor. In order to perform high-bandwidth measurements of the sensor impedance, we make use of RF-reflectometry, where the SHT is part of a resonant LCR-circuit further consisting of an off-chip superconducting resonator together with the parasitic device capacitance. We apply a microwave signal to the tank circuit and measure the amplitude of the signal reflected by the LCR-circuit (see Figure \ref{fig:1}a). The amplitude of the reflected signal $|S_{12}|$ depends on the matching of the tank circuit impedance with the measurement setup and is therefore modulated by a change in the charge sensor impedance caused by the movement of a nearby charge.

\begin{figure*}
    \centering
	\includegraphics{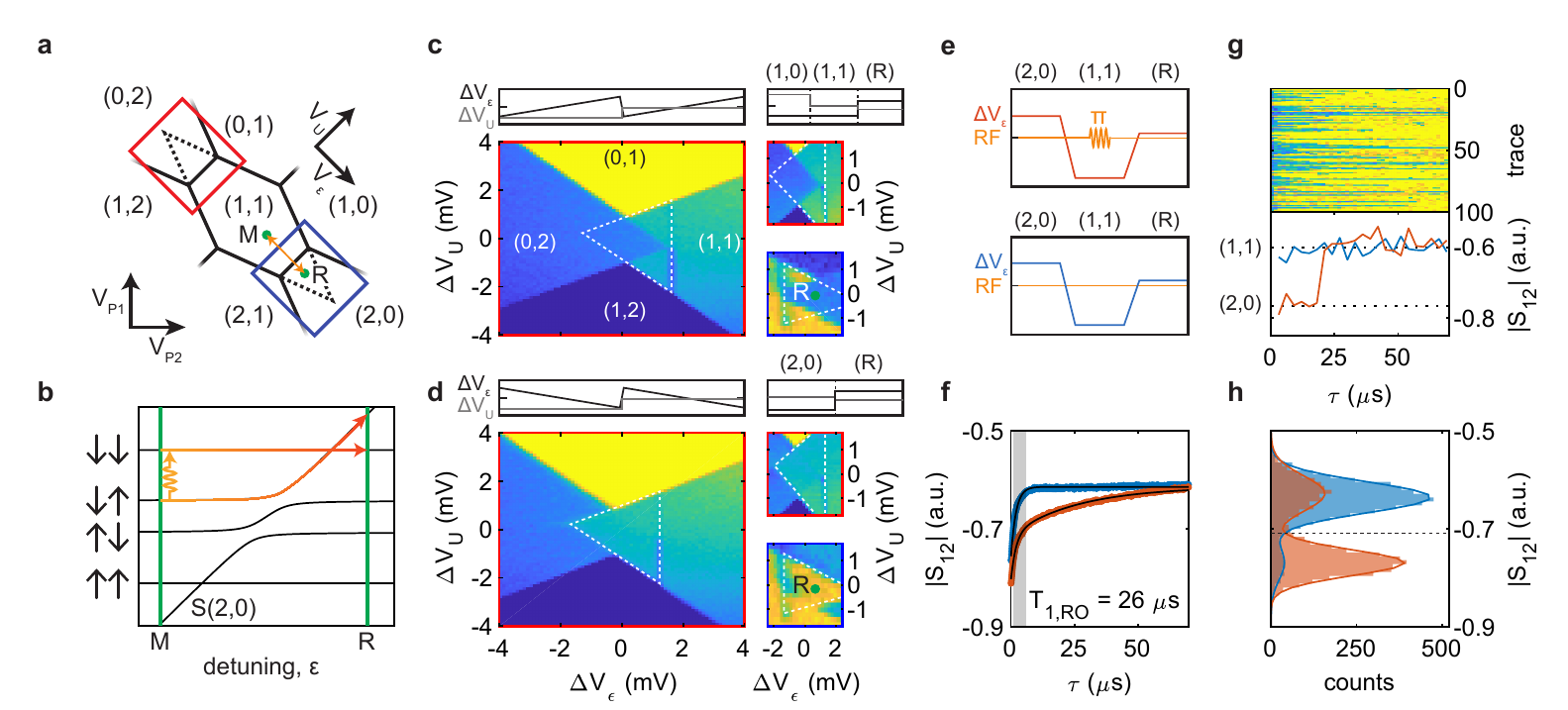}%
	\caption{
	\textbf{Single shot spin readout of a single hole}
	\textbf{a} Schematic of a typical hole charge stability diagram with both possible regions of readout indicated in blue and red respectively. The typical manipulation (M) and readout (R) points are indicated in green.
	\textbf{b} Two-hole energy diagram, with the five lowest lying energy states around the (1,1)-S(2,0) anticrossing. By adiabatically pulsing from M to R, the system will end up either in the (1,1) or (2,0) charge state depending on the spin state of the two holes, resulting in a spin-to-charge conversion.
	\textbf{c,d} Colour map of the sensor response as a function of the applied gate voltages $U$ and $\epsilon$. In the large panels on the left, we linearly sweep $\epsilon$ and step $U$. Sweeping from the (0,2) into the (1,1) region (\textbf{c}), moves the holes into the same quantum dot, while sweeping in the opposite direction results in spin blockade (\textbf{d}). The smaller panels on the right show the same effect, but now using a three level voltage pulse to load a random spin and integrating the signal for 10 $\mu$s at each pixel. The readout point used in the further experiments is indicated by R.
	\textbf{e} Schematic illustrating the three-level pulses used in panels f-h, indicating the detuning voltage $\Delta V_\epsilon$ in blue and red and the RF-pulses in orange. We prepare the system in the $\ket{\downarrow\uparrow}$ state by adiabatically sweeping across the anticrossing and apply an additional $\pi$-pulse to Q1 to initialize the system in the $\ket{\uparrow\uparrow}$ triplet state. Typically adiabatic ramp times of $t_\text{ramp}\approx 1~\mu$s are used.
	\textbf{f} The averaged charge sensor response as a function of measurement time $\tau$ at R shows the spin relaxation decay. Both lines consist of 5000 averaged single shot traces. The gray shaded area indicates the integration window for the single shot detection.
	\textbf{g} A sample of 100 single-shot traces (top), averaging for 3~$\mu$s per data point, with $\tau=0$ equals the start of the readout phase. Two single traces are plotted in the bottom panel, where the blue (red) trace corresponds to the system being prepared in the $\ket{\downarrow\uparrow}$ ($\ket{\uparrow\uparrow}$) state. The dashed lines correspond to the sensor response to the different charge states.
	\textbf{h} Histogram of 5000 single shot traces, integrating over 5.5~$\mu$s as indicated in panel \textbf{f}. Again, the blue (red) histogram corresponds to an initialisation in the $\ket{\downarrow\uparrow}$ ($\ket{\uparrow\uparrow}$) state. For both histograms, two clear peaks are visible, corresponding to the singlet and triplet readout respectively, with the weight shifting between the peaks depending on the state preparation. The dashed line corresponds to the optimized threshold for readout.
	}
	\label{fig:2}
\end{figure*}

We make use of the RF sensor to map out the charge stability diagram of the double quantum dot system defined by plunger gates P1 and P2. The tunnel coupling of the quantum dots to their reservoirs, as well as the interdot tunnel coupling can be tuned by gates RB1, RB2 an B12 respectively. Next, we tune the device to the single hole regime for both quantum dots (Fig. \ref{fig:1}b and Supp. Fig. 1, where ($N_1$,$N_2$) indicates the charge occupation, with $N_1$ ($N_2$) the hole number in Q1 (Q2), see Methods. In order to perform readout of the spin states, we make use of Pauli spin blockade (PSB), which is expected to be observed both at the (1,1)-(0,2) and (1,1)-(2,0) charge transitions. We define the virtual gates \cite{hensgens_quantum_2017} detuning $V_\epsilon$ and energy $V_U$ (see Fig. \ref{fig:2}a and Methods) and sweep across the (1,1)-(2,0) and (1,1)-(0,2) transitions in this gate space. As a result of its triplet character, the $\ket{\downarrow\downarrow}$ state has a negligible coupling to the S(2,0) or S(0,2) singlet charge states (Fig. \ref{fig:2}b). When pulsing across the (1,1)-S(2,0) or (1,1)-S(0,2) anti-crossings, the hole is not allowed to tunnel into the other quantum dot when the system is in the $\ket{\downarrow\downarrow}$ ground state. However, when the system resides in the singlet-like lower antiparallel spin state (in this case $\ket{\downarrow\uparrow}$), the hole can tunnel into the other quantum dot, therefore leaving the system in a (0,2) or (2,0) charge state. This results in a spin-to-charge conversion, which in turn can be picked up in the reflectometry signal from the SHT.

Indeed, we find that by sweeping the detuning across the interdot transition from the (1,1) into the (0,2) charge region, tunneling is blocked (Fig. \ref{fig:2}d) up to the reservoir transitions (indicated in white), when the system is initialised in the $\ket{\downarrow\downarrow}$ state. In this case we rely on the fast diabatic return sweep combined with fast spin relaxation to prepare the system in the blocking $\ket{\downarrow\downarrow}$ state. When we inverse the sweeping direction, the system remains in the (0,2) charge states at the same values of $V_\epsilon$ and $V_U$ (Fig. \ref{fig:2}c). After optimizing the different tunnel rates in the device, we confirm the Pauli spin blockade at both the (1,1)-(2,0) and (1,1)-(0,2) anticrossings by loading a random spin before performing the readout, thereby not relying on a relaxation process for the initialisation (small panels of Fig. \ref{fig:2}c,d). The diamond-shaped window of differential signal allows for a singlet/triplet readout of the system spin state and we select readout point R (see Supp. Fig. 2). We note that the interdot transition line is shifted slightly towards positive detuning with respect to the reservoir transition lines. This is the direct result of a small thermal voltage present across the device ohmics, resulting in the unusual diamond-shaped spin readout window, but not limiting the readout. As holes in germanium do not have any valley states, the T(2,0) state is expected to be defined by the next quantum dot orbital. By increasing the bias voltage across the two quantum dots, we shift the interdot transition line. At large enough bias, the Pauli spin blockade window is capped as a result of the T(2,0) state being available in energy and from this we extract an excited state energy of $E_{ST} = 0.85$ meV, using a lever arm of $\alpha_\epsilon=0.21$ as extracted from polarisation line measurements (Supp. Fig. 3).

To coherently control the qubits, we implement a three-level voltage pulsing scheme (Fig. \ref{fig:2}e) and operate at an external magnetic field of $B=0.67$ T. We initialise the system by pulsing deep into the (2,0) region, where the spins quickly relax into the (2,0) singlet state. Next, we ramp adiabatically into the (1,1) region, preparing the system into the $\ket{\downarrow\uparrow}$ state. At this point (M) we perform the qubit operations by applying microwave pulses to gate P1, taking advantage of the SOC-mediated EDSR. Rotating Q1 (Q2) will bring the system into the $\ket{\uparrow\uparrow}$ ($\ket{\downarrow\downarrow}$) state. Finally, the spin-state is read out by pulsing adiabatically into the readout window. Only the $\ket{\downarrow\uparrow}$ will allow a direct tunneling into the (1,1) charge state, where tunneling is blocked for all other states due to PSB.

Fig. \ref{fig:2}f displays the decay of the two-qubit spin-state throughout the readout period, with and without a $\pi$-pulse applied to Q1. When no pulse is applied and the system is prepared in the $\ket{\downarrow\uparrow}$ state, a fast transition into the (1,1)-charge state, corresponding to a sensor signal of $|S_{12}|\approx-0.6$ can be observed. The remaining decay ($T_\text{decay}=2~\mu$s) in this case can be attributed to the response of the SHT-signal to the voltage pulses on the gates. However, when the system is prepared in the $\ket{\uparrow\uparrow}$ state by applying a $\pi$-pulse to Q1, a significantly slower relaxation into the (1,1) state is observed, due to the slow T$_+$(1,1)-S(2,0) relaxation. By fitting a double exponential decay, accounting for the SHT response, we extract a spin relaxation at the readout point of $T_{1,\text{RO}}=26~\mu$s. A sample of 100 single-shot traces is plotted in Fig. \ref{fig:2}g, together with two individual traces using a post-processing integration time of $3~\mu$s. A clear distinction of the (1,1) and (2,0) charge states can be observed from the sensor response. To determine the spin state of the qubits, we perform a threshold detection of the single-shot signal integrated from $\tau_0=1.0~\mu$s up to $\tau_\text{meas}=6.0~\mu$s for maximised visibility, discarding the initial stabilisation of the SHT and optimizing between the charge discrimination and spin relaxation. A histogram of 5000 single-shot events illustrates the clear distinction between the singlet ($|S_{12}|>-0.72$) and the triplet ($|S_{12}|<-0.72$) spin state readout. We find a spin readout visibility of $v=56\%$ as obtained from the difference in spin-up fraction between the two prepared states. A large part of this reduced visibility is caused by relaxation of the blocked triplet state during the measurement, expected to amount to a signal reduction of $P_\text{relax}=1-e^{-\tau_\text{meas}/T_{1,\text{RO}}}=0.21$. This gives good prospects for increasing the readout fidelity by optimising the spin relaxation, for instance by optimizing the reservoir tunnel rates and moving to latched PSB readout mechanisms \cite{studenikin_enhanced_2012,harvey-collard_high-fidelity_2018}. Alternatively, by using high-Q on-chip resonators \cite{zheng_rapid_2019} the signal-to-noise ratio could be significantly improved, thereby lowering the required integration time and reducing the effective relaxation. The remaining triplet fraction of $0.11$ that can be observed for the readout of the $\ket{\downarrow\uparrow}$ state could be attributed to an anadiabaticity of the pulsing or a small coupling between the T(1,1) and S(2,0) states as mediated by the SOC, which could be mitigated by further optimizing the readout pulse sequence.

\begin{figure}
    \centering
	\includegraphics[]{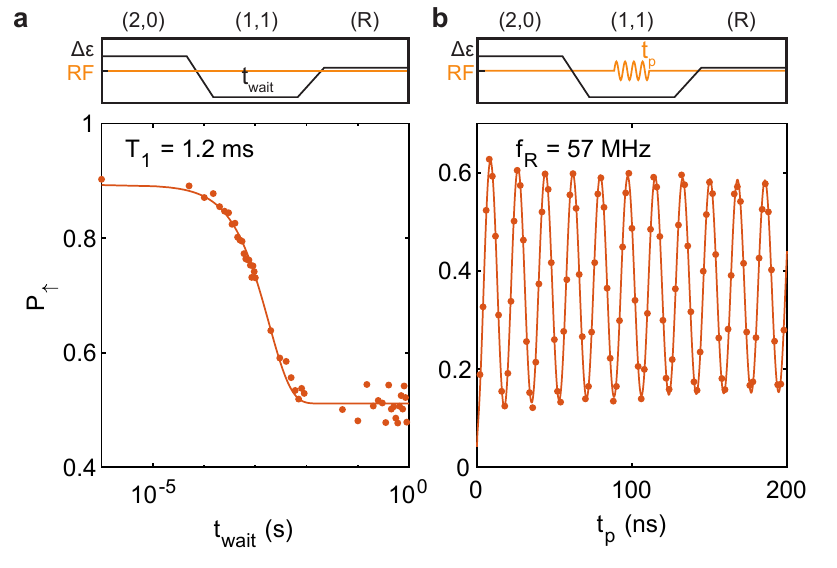}%
	\caption{
	\textbf{Spin relaxation and coherent driving of a single hole.}
	\textbf{a} The system is initialized in the $\ket{\downarrow\uparrow}$ state after which the qubits idle at the measurement point. The spin-up fraction $P_\uparrow$ of Q2 is measured as a function of waiting time $t_\text{wait}$ and shows a typical $T_1$-decay with $T_1=1.2$~ms.
	\textbf{b} Driving of the single hole qubit Q2 shows coherent oscillations in $P_\uparrow$ as a function of the microwave pulse length $t_\text{p}$.
	\label{fig:3}}
\end{figure}

Now we probe the single spin relaxation time by initialising the system in the $\ket{\downarrow\uparrow}$ state and letting the system evolve at a detuning voltage $\Delta V_\epsilon=-7$~mV from the (1,1)-(2,0) anticrossing. Fig \ref{fig:3}a shows the spin-up fraction as a function of the waiting time $t_\text{wait}$, from which a single spin relaxation time of $T_{1,\text{Q1}}=1.2$~ms can be extracted. This is substantially longer than reported before in planar germanium heterostructures \cite{hendrickx_fast_2019}, most likely as a result of the more isolated single hole spins as compared to the transport measurements with high reservoir couplings, and is now also longer than all relevant time scales for qubit operation. This relaxation time also compares favourably to results obtained for holes in Ge hutwires \cite{vukusic_single-shot_2018} and other hole spins \cite{bogan_single_2019, gerardot_optical_2008, hu_hole_2012}. 

\begin{figure}
    \centering
	\includegraphics[width=\columnwidth]{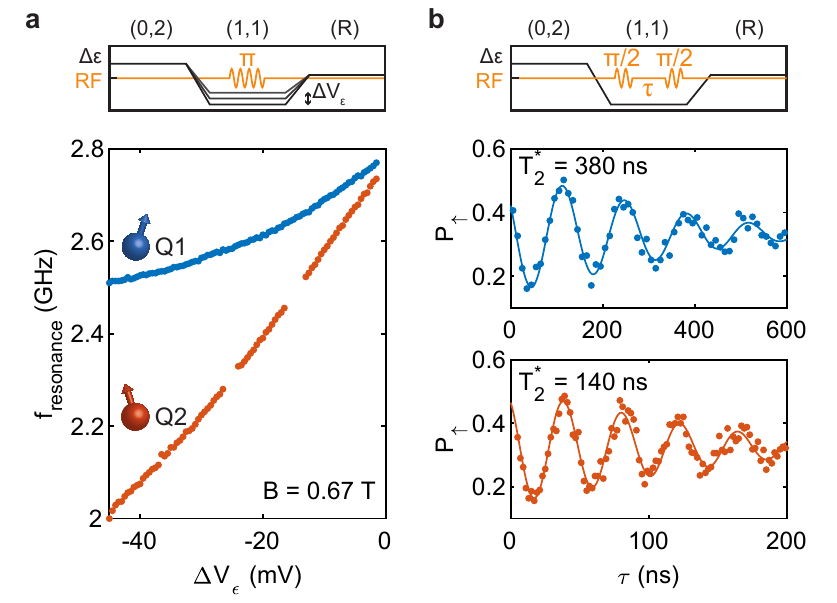}%
	\caption{
	\textbf{Electric g-factor modulation and phase coherence of the qubit resonances.}
	\textbf{a} The resonance frequency of both qubits shows a strong modulation as a function of the detuning voltage $\Delta V_\epsilon$.
	\textbf{b} We perform a Ramsey experiment on both qubits to probe the phase coherence times, with $T_{2,Q1}^*=330$ ns and $T_{2,Q2}^*=130$ ns. The comparatively shorter phase coherence can be attributed to the strong dependence of $f_\text{resonance}$ to electric fields, coupling charge noise to the spin state.}
	\label{fig:4}
\end{figure}

To demonstrate coherent control of the single hole, we modulate the length of the driving microwave pulse and measure the spin-up fraction (Fig. \ref{fig:3}a). A clear sinusoidal Rabi oscillation can be observed, with a Rabi frequency of $f_R=57$~MHz (coherent operation of Q2 in Supp. Fig. 4). We probe the phase coherence of both qubits by performing a Ramsey sequence in which we apply two $\pi/2$ pulses, separated by a time $\tau$ in which we let the qubit freely evolve and precess at a frequency offset of $\Delta f=7.4$~MHz and $\Delta f=23.7$~MHz respectively. In Fig. \ref{fig:4} a,b the Ramsey decay for Q1 and Q2 are plotted and we extract coherence times of $T_2^*=330$~ns and $T_2^*=130$~ns respectively. These coherence times are of comparable order, but slightly lower than previously reported numbers in the same heterostructere for a many-hole quantum dot \cite{hendrickx_fast_2019}. In order to explain the origin of this, we measure the resonance frequency of both qubits as a function of the detuning voltage $\Delta V_\epsilon$. We find a very strong dependence of the resonance frequency of both qubits on the detuning voltage over the entire range of voltages measured, with the $g$-factor varying between $g_\text{Q1}=0.27-0.3$ and $g_\text{Q2}=0.21-0.29$ for Q1 and Q2 respectively. This strong electric field dependence of the resonance frequency will increase the coupling of charge noise to the qubit spin states, which in turn will reduce phase coherence \cite{hendrickx_fast_2019}. The ratio in local slopes of the resonance frequency $\delta f_\text{Q1}/\delta f_\text{Q2}=2$ is similar to the ratio in phase coherence of both qubits $T_{2,Q1}^*/T_{2,Q2}^*=2.5$, consistent with charge-noise limited coherence. The strong modulation of the qubit resonance frequency by electric field can be explained from the strong SOC present \cite{maier_tunable_2013}, as also suggested by the Rabi frequency changing as a function of detuning voltage (see Supp. Fig. 5). Although the strong $g$-factor modulation seems mainly a cause of decoherence in this case, careful optimisation of the electric field landscape could render a situation in which the qubit Zeeman splitting is well controllable, while maintaining a zero local slope for high coherence \cite{culcer_transport_2019}.

The demonstration that single hole spins can be coherently controlled and read out in single-shot mode, together with the spin relaxation times $T_1>1$ ms, defines planar germanium as a mature quantum platform. These aspects are demonstrated on a two-dimensional quantum dot array, further highlighting the advancement of germanium quantum dots. Moreover, controlling a single hole spin represents an important step towards reproducible quantum hardware for scalable quantum information processing.

\bibliography{spinqubit}
\section*{Acknowledgements}
We thank R.N. Schouten and M. Tiggelman for their assistance with the measurement electronics. We acknowledge support through a FOM Projectruimte and through a Vidi programme, both associated with the Netherlands Organisation for Scientific Research (NWO).


\section*{Author Information}
The authors declare no competing financial interests. Correspondence should be addressed to M.V. (M.Veldhorst@tudelft.nl).


\section*{Methods}
\small
\subsection{Fabrication process}
We grow Ge/SiGe heterostructures on an $n$-type Si(001) substrate, using an Epsilon 2000 (ASMI) RP-CVD reactor, as further discussed in Ref. \cite{sammak_shallow_2019}. Ohmic contacts are defined by electron beam lithography, electron beam evaporation and lift-off of a 20-nm-thick Al layer. Electrostatic gates consist of a Ti/Pd layer with a thickness of 20 and 40 nm respectively for the barrier and plunger gate layer. Both layers are separated from the substrate and each other by 10 nm of ALD-grown Al$_2$O$_3$.

\subsection{Experimental setup}
We use a Bluefors dry dilution refridgerator with a base temperature of $T_\text{bath}\approx10$ mK to perform the measurements. Battery-powered voltage sources are used to supply DC-voltages on the gates. Additionally, AC-voltages generated by a Tektronix AWG5014C arbitrary waveform generator can be supplied to the gates through a bias-tee with a cut-off frequency of $\approx 10$ Hz. Similarly, we can also apply a microwave signal generated by a Keysight PSG8267D vector source to gate P1 for qubit driving. Driving both qubits at the same power on gate P1, we observe significantly faster Rabi oscillations in Q2. From this we assume Q2 to be located under P1, and Q1 under P2, in correspondence with the trend in Rabi frequencies observed in a previous work \cite{hendrickx_fast_2019}.

We use an in-house built RF generator to supply the reflectometry signal. The signal is attenuated by 84 dB and applied to one of the sensor Ohmics via a Mini-Circuits ZEDC-15-2B directional coupler. The reflected signal is amplified by a Caltech CIRLF3 SiGe-amplifier at the 4K-stage of our fridge and an in-house built RF-amplifier at room temperature, and demodulated to give a measure of $|S_{12}|$.

\subsection{Virtual gates}
In order to allow independent control over the detuning and energy of the quantum dots more easily, we define the virtual gate axes of $V_\varepsilon=V_\text{P2}-0.5V_\text{P1}$ and $V_U=0.5V_\text{P2}+V_\text{P1}$.

\end{document}